\documentclass[a4paper,UKenglish,cleveref, autoref, thm-restate,numberwithinsect,nameinlink]{lipics-v2021}

\title{Lower bounds for dominating set reconfiguration on sparse (directed) graphs.} %TODO Please add
\author{Jona Dirks}{University Clermont Auvergne, France}{jona.dirks@uca.fr}{https://orcid.org/0009-0002-9580-9461}{}

\newtheorem{Gadget}[theorem]{Construction}

\author{Alexandre Vigny}{University Clermont Auvergne, France}{alexandre.vigny@uca.fr}{https://orcid.org/0000-0002-4298-8876}{}
 
\authorrunning{J. Dirks and A. Vigny} %TODO mandatory. First: Use abbreviated first/middle names. Second (only in severe cases): Use first author plus 'et al.'

\Copyright{Jona Dirks and Alexandre Vigny} %TODO mandatory, please use full first names. LIPIcs license is "CC-BY";  http://creativecommons.org/licenses/by/3.0/


 %TODO mandatory; please add comma-separated list of keywords

\category{} %optional, e.g. invited paper

\relatedversion{} %optional, e.g. full version hosted on arXiv, HAL, or other respository/website
\nolinenumbers %uncomment to disable line numbering

\ArticleNo{1}
\usepackage{booktabs}

\usepackage[]{blindtext}
\usepackage{complexity}
\usepackage{algorithm2e}
\usepackage{subcaption}
\usepackage{subfiles}
\usepackage{wasysym}
\usepackage{lineno} % Line numbering package

\usepackage{amsthm}
\usepackage{mathrsfs}
\usepackage{mathtools}
\usepackage{todonotes}
\usepackage{float}

\newcounter{prop}

\newtheorem{property}[prop]{Property}

\newcommand{\tw}{\ensuremath{\textrm{tw}}}
\newcommand{\pw}{\ensuremath{\textrm{pw}}}
\newcommand{\dfvs}{\ensuremath{\textrm{dfvs}}}

\newcommand{\interval}[2]{\ensuremath{[#1,#2]}}

\usepackage{tikz}
\usetikzlibrary{decorations.pathmorphing, calc}
\usetikzlibrary{arrows.meta}
\pgfdeclarelayer{background}
\pgfdeclarelayer{foreground}
\pgfsetlayers{background,main,foreground}

\newcommand{\bag}{\mathsf{bag}}

\newcommand{\isdts}{ISR-DTS\xspace}
\newcommand{\ists}{ISR-TS\xspace}

\newcommand{\dsts}{DSR-TS\xspace}
\newcommand{\dsdts}{DSR-DTS\xspace}
\newcommand{\rbdsdts}{ReBuDSR-DTS\xspace}
\newcommand{\rbdsts}{ReBuDSR-TS\xspace}

\newcommand{\validator}{\textsf{validator}}

\renewcommand{\FPT}{\textup{FPT}\xspace}
\renewcommand{\PSPACE}{\textup{PSPACE}\xspace}

\usepackage{contour}
\usepackage{ulem}

\contourlength{0.8pt}

\renewcommand{\underline}[1]{%
	\uline{\phantom{#1}}%
	\llap{\contour{white}{#1}}%
}
\nolinenumbers %uncomment to disable line numbering
\begin{document}
    
    \maketitle
    \begin{abstract}
        In a graph, a vertex dominates itself and its neighbors, and a dominating set is a set of vertices that together dominate the entire graph. Given a graph and two dominating sets of equal size $k$, the {\em Dominating Set Reconfiguration with Token sliding} (\dsts) problem asks whether one can, by iteratively replacing a vertex by an adjacent one, transform the first set into the second one, while ensuring that every set during the reconfiguration process is a dominating set.

        The token jumping variant, where a vertex can be replaced by a non-adjacent one, is known to be efficiently solvable on many graph classes such as planar, bounded treewidth, and the very broad notion of nowhere-dense classes of graphs. 
        Alternatively, some algorithms also exist for the reconfiguration of independent sets in the token sliding paradigm for graph classes with bounded degree or large girth.

        We show that \dsts is W[2]-hard when parameterized $k$, the pathwidth of the instance, and the iteration of the reconfiguration sequence 
        % (a parameter recently introduced in [Driks, Vigny, MFCS 2025]). 
        (a recently introduced parameter). 
        This is a setting where both the token jumping and the independent set variants are fixed parameter tractable. Not restricting the iteration yields W[2] hardness already on graphs with treewidth 9 and pathwidth 13.
        
        % \newJona{In the directed variant (DSR-DTS), we are only allowed to replace a vertex with an out-neighbor. We show that DSR-DTS is W[2]-hard in DAGs when parameterized by the size of the sets. Even if the length of a directed path is at most 3 (independent set reconfiguration is again FPT in this setting) and NP-hard in DAGs even when the treewidth is 5.}

        In the directed variant (DSR-DTS), we are only allowed to replace a vertex with an out-neighbor. We show that DSR-DTS is NP-hard on DAGs of treewidth 5 and W[2]-hard for both the case of DAGs of depth 3 parameterized by $k$, and the case of DAGs when parameterized by $k$ and the pathwidth of the instance (independent set reconfiguration is again FPT in both settings).
    \end{abstract}

    \section{Introduction}
    %!TEX root = ../paper.tex
% {
% \color{red}
% TODO:
% \begin{itemize}
% 		\item more gentle introduction into the topic.
% 		\item Citing the relevant Papers:
% 		\begin{itemize}
% 			\item \cite{Gima2024} potentially we can build in metha-theorems into the motivation (also written by one of the chairs)
% 			\item \cite{Lokshtanov2022} maybe somewhere mention connected dominating set (also written by one of the chairs) (There are why more papers by Lokshtanov potentially some are relevant)
% 			\item \cite{Agrawal2025} (also written by one of the chairs)  there are multiple version did I get the right one?
			
% 		\end{itemize}
% \end{itemize}

% }

Many real world situations either deal with movement or changing states in general. These problems can be seen as reconfiguration problems where we are given two solutions to some problem and are asked to change configurations from the first to the second, step by step according to some rule and maintaining a solution all along.

For problems defined over subset of vertices (such as Dominating Set or Independent Set), we can treat the solution as token traveling over the graph. Two main relocation rules are studied in the literature: {\em Token Sliding} where we are allowed to move a token to one of its neighbor (i.e.~to slide the token along an edge), and {\em Token Jumping}, where we are allowed to move a token from a vertex to any other vertex~\cite{Bousquet2024}. Somewhat recently a variation was introduced for directed graphs: for {\em Directed Token Sliding}, we are allowed to slide a token only along an edge in the direction of the edge \cite{ito2022}.
	
% Recently, it was shown that \isdts is in \FPT\xspace for DAGs with when parameterized by $k+\tw$. 

% A reconfiguration problem asks, given two solutions to a combinatorial problem, whether
% one can change the first solution in a step-by-step manner in order to reach the second one;
% while ensuring that the transformed set remains a solution throughout the reconfiguration
% procedure. For these problems, there are two main reconfiguration paradigms: {\em token sliding}, and {\em token jumping}. In both paradigms, we view the set to be reconfigured as several tokens placed on the vertices of the graph. Token jumping allows, in each step to move one token to be placed to any other vertex of the graph, while token sliding only allows to move a token to an adjacent vertex (to slide along an edge of the graph).

Specific attention has been given to independent set and (connected) dominating set reconfiguration. We refer the readers to the survey \cite{Bousquet2024} for an overview of the known (parametrized) algorithms for both token jumping and sliding paradigms. 
In a nutshell, reconfiguration (for both problems) in the token jumping model are  \PSPACE-complete, and remains intractable when parameterized by the size  $k$ of the sets to be reconfigured, or the length of reconfiguration sequence~\cite{BodlaenderGS21,MouawadN0SS17}.
When looking at structural parameters of the given graphs however, independent set and (connected) dominating set become \FPT on classes of graphs with bounded treewidth, bounded degree, planar graphs and even nowhere dense graph classes~\cite{Bousquet2024,LokshtanovM19,Siebertz18}.

For token sliding however much less is known. It is only known that independent set reconfiguration (\ists) is \FPT on trees~\cite{DemaineDFHIOOUY14}, graphs with bounded degree~\cite{Bartier2023}, or girth 5~\cite{BartierBHMS24}, but hard on split graphs~\cite{BonamyB17}.
The main pressing open question is whether independent set reconfiguration is \FPT (when parametrized by $k$) on graph classes with bounded treewidth or minor-free graphs~\cite[Question 5]{Bousquet2024}. Alternatively, some studies looked at the length of the reconfiguration sequence as an additional parameter~\cite{Agrawal2025,Gima2024,Lokshtanov2022,Wrochna2018}

Quite surprising, nothing non-trivial is known for dominating set reconfiguration under token sliding (\dsts)~\cite[Question 10]{Bousquet2024}.
In this paper we show that the lack of known algorithm in this setting is due to the fact that \dsts is W[2]-hard parametrized by $k$ already on classes with treewidth 10 and pathwidth 13, see \cref{New2}.

\medskip\noindent{\bf Recent developments.}
A similar result has just been independently and simultaneously proved by \cite{Bousquet2025}. One difference is that the lower bound they provide if much higher than ours. They show XL-completness while we only show W[2]-hardness. The techniques are different, while they introduce a new setting they call {\em tape reconfiguration}, we make reduction from dominating set. Note that the bounds we obtained are smaller than theirs e.g.~hardness for classes with treewidth 10 and pathwidth 13 in our \cref{New2} compared with hardness for treewidth at most 12 and pathwidth at most 18 in \cite[Theorem 2]{Bousquet2025}.

More importantly, we study more restricted cases where \ists is known to be \FPT : that is by restricting the iteration, a parameter introduced in \cite{Driks2025}. We therefore describe for the first time a setting where \ists is \FPT but \dsts is not (assuming W[2]$\neq$ \FPT).
Furthermore, we explore the recent variant of {\em Directed Token Sliding} and also show hardness in places where \ists is \FPT showcasing once more the stark differences between the two problems.

This is somewhat counterintuitive because while independent set and dominating set problems are not is the same parametrized complexity classes (\W[1]-hard and \W[2]-hard respectively) they are generally either both tractable or both hard \cite{FabianskiPST19}. 

\medskip\noindent{\bf The directed setting.}
Introduced by \cite{ito2022} in 2022, the {\em Directed Token Sliding} paradigm works on directed graphs and ask the reconfiguration move to follow the direction of the edges.
When introducing the independent set reconfiguration under directed token sliding (\isdts) problem they show that it is already W[1]-hard on DAGs \cite[Theorem 2]{ito2022}.

However, it is proved in \cite[Theorem 1.4]{Driks2025} that \ists is \FPT on DAGs, when parametrized by $k$ and the (undirected) treewidth of the instance.
It would then fell natural that the dominating set variant \dsdts is also tractable in this setting, but we show that this is not the case (see \cref{theo:dsdtsdaghard}). This paper can be seen as a testimonial to how hard Dominating Set Reconfiguration by Token Sliding and its variation actually are.

We also look into the depth of the DAGs as a tractability parameter and show a dichotomy between \dsdts being polynomial on Dags of depth 2 but already NP-hard on DAGs of depth 3 (see \cref{theo:dagdepth2,theo:dagdepth3}).

\medskip\noindent{\bf Techniques and structure of the paper}
We work with a red-blue variant of dominating set reconfiguration and later reduce \dsts to this problem on various settings.

In \cref{sec:depth} we look at the depth of DAGs as a possible tractability parameter. In \cref{sec:undirected-reconfig} we study the undirected reconfiguration of dominating set. First it's red-blue variant, where the token moves are more constraint and hardness is easier to show, and then the consequence for the uncolored version.
After that, in \cref{sec:directedReconfig} we move to the directed setting, and finally in \cref{sec:np-best-bounds} we provide better lower bounds concerning NP-hardness.

    \section{Preliminaries}
    \paragraph*{Graphs} We use standard graph notation. Unless stated otherwise, all graphs are directed and loop-free.  We denote the set of all directed cycles in a graph $G$ with $C(G)$. \emph{Directed Acyclic Graphs} (DAGs) are graphs with $C(G) = \emptyset$. We can layer a DAG, by partitioning the set of vertices. Vertices in {\em layer} $1$ are all the vertices with no incoming edges, and inductively the vertices in layer $i$ are those with in-neighbors only in layers at most $i-1$.
The \emph{depth} of a DAG is the highest index of a layer.
This is equivalent to the \emph{length} of the longest directed path, defined as the number of vertices on it.

The {\em open} neighborhood $N(v)$ of a vertex $v$ is the set of every adjacent vertices. The {\em closed} neighborhood $N[v]$ is the open neighborhood together with the vertex itself. This is adapted to sets by taking the union of the (closed or open) neighborhood of every vertex in the given set.

\paragraph*{Dominating Set and its Variations}
A Dominating set is a subset of vertices $S\subseteq V(G)$ of an graph $G$ such that $N[S] = V(G)$.

In Red Blue Dominating Set we partition the vertices into red and blue vertices. We are asked to find a subset of blue vertices that dominate all red vertices.

We define both Dominating Set and Red Blue Dominating Set for directed graphs disregarding the edge orientation. 
That is a (Red Blue) Dominating Set is exactly a (Red Blue) Dominating Set in the underlying undirected graph.
One could ask for a Dominating Set in a directed graph to out-dominate the entire graph (or every red vertex). In some of our results for directed graphs (see~\cref{theo:rbdsdtsDAGW2,theo:rbdsdtsW2hardfixtw}) this choice has no impact has the dominating sets we build in our proof have both properties. However, in~\cref{sec:depth} our construction is tailored to dominating sets for the undirected underlying graph.

\paragraph*{Parameterized Complexity}
Problems with parameter $k$ are \emph{fixed parameter tractable} (\FPT) if an algorithm solving it in time $f(k)\cdot n^c$ exists, for a computable function $f$ and constant $c$.

 Problems that are \emph{$W[i]$-hard} (for every integer $i$) are considered not to be efficiently solvable. Dominating Set is known to be $W[1]$-hard. Mirroring \NP-hardness, we can transfer W[2]-hardness using \FPT-reductions. That is a function. computable \FPT-time, that for every instance returns an equivalent instance with the parameter being restricted by $g(k)$, for some function $g$.
 
 \paragraph*{Reconfiguration} Let $T=\{t_1,t_2,\dots,t_k\}$ be a set of tokens. For two dominating sets $S,D$ in a graph $G$, a \emph{reconfiguration sequence} $(\alpha_i)_{i\in I}$ is a sequence of configurations $\alpha_i :T \to V(G)$, such that
 \begin{enumerate}
 	\item $\alpha_1(T) = S$ and $\alpha_{|I|}(T) = D$,
 	\item for every $i$, $\alpha_i(D)$ is a dominating set of $G$, and
 	\item for every $i$ there exists only one token $t$ such that $\alpha_{i-1}(t) \neq \alpha_i(t)$.
 \end{enumerate}
 These rules define the token jumping variant of Dominating Set Reconfiguration. We will look at two restrictions regarding token relocation of this problem throughout this paper. For \emph{Dominating Set Reconfiguration under Token Sliding (\dsts)} we require that for each $i \in I, i>1$ if $\alpha_{i-1}(t) \neq \alpha_i(t)$, then the graph must contain the undirected edge $\alpha_{i-1}(t)\alpha_i(t)$. 
 For \emph{Dominating Set Reconfiguration under Directed Token Sliding (\dsdts)} we require that if $\alpha_{i-1}(t) \neq \alpha_i(t)$, then the graph contains the directed edge $\alpha_{i-1}(t)\alpha_i(t)$.
 
 For our proofs we will use Red Blue Dominating Set Reconfiguration (abbreviate \rbdsts and \rbdsdts depending on the setting). Here we require that~$S,D$ and every $\alpha_i(T)$ is a red blue dominating set. Note that this means that no token can be placed on a red vertex throughout the reconfiguration sequence. We can also assume without loss of generality that in the directed setting edges are always directed from blue to red vertices.\
 
 We use $k$ for the size of the sets $S$, $D$, $T$ and thus the size of every configuration.
 
The iteration measures how often a token visits a vertex maximal. It is defined as follows.
\begin{definition}[Iteration, from \cite{Driks2025} Definition 7.1]
	Given a reconfiguration sequence $(\alpha_i)_{i\in I}$, a token $t$ and a vertex $v$, the {\em iteration} of $(v,t)$ is the number of times the token enters $v$ (plus one if $t$ starts in~$v$).
	The {\em iteration}  $\iota$ of a reconfiguration sequence $(\alpha_i)_{i\in I}$ is the maximum iteration of all pair~$(v,t)$:\\	
	$\iota := \max\limits_{(v,t)\in V\times T}|\{0~|~\alpha_0(t) = v\} \cup \{i\in I~|~0<i\textrm{ and } \alpha_{i-1}(t)\neq\alpha_{i}(t) = v\}|$.
	
\end{definition}

 %\todo{Missing the parameter $\iota$}
 %\todo{Potentially missing \rbdsdts}
 %\todo{defin \dsts \dsdts \rbdsdts, and configuration vs reconfiguration sequence}
 %\todo{$k$ always number of tokens}

 \paragraph*{Tree and Pathwidth}
 For an undirected graph $G$, a \emph{tree-decomposition} is a pair $\mathcal{T} = (T,\bag)$, where $T$ is a rooted tree and $\bag : V(T) \to 2^{V(G)}$ assigns each node $x$ of $T$ its \emph{bag} $\bag(x)$ such that:
 \begin{enumerate}
 	\item For every vertex $v$ the subtree induced by all bags that contain $v$ is connected and non-empty.
 	\item For every edge $uv \in E(G)$ there exists a node $x \in V(T)$ with $u,v\in \bag(v)$.
 \end{enumerate}
 
 The \emph{treewidth} is defined as the size of the larges bag plus one. \emph{Pathwidth} and \emph{path-decompositions} are defined analogously, with $T$ being a path.
 
%  For a graph partitioned into red and blue vertices, we call a path decomposition \emph{well-structured} if for each blue connected component $C_i$ there exists a subpath $P_i$ of the path-decomposition such that its bags contain all vertices of $C_i$ and such that all paths  $P_i$ are pairwise vertex disjoint.\alex{to remove, this is almost property 2}
 
\paragraph*{Intervals and Modular Arithmetic}
Let $i,j$ and $\ell$ be integers.
We denote the interval of integers between $i$ and $j$ with $\interval{i}{j}$. Throughout this paper we will use modular arithmetic. We say that $i \equiv j \mod \ell$ if $i$ and~$j$ have the same reminder when divided by $\ell$. The representative of the equivalent class is the reminder itself denoted $[j \mod \ell]$.
    \section{DAGs of low depth}\label{sec:depth}
    \begin{figure}
	\includegraphics[width=\linewidth]{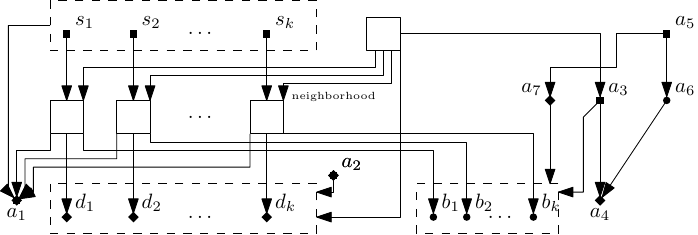}
	\caption{DAG of depth 3 resulting from the reduction. Squares are start positions, diamonds destination and circle neither. Note that $a_1$ and $a_2$ are both squares and diamonds.}
\end{figure}
%\textcolor{red}{We dont see the terminals for a 3457. Maybee to notation like the square and diamonds of figure 2?}}

\begin{theorem}
	\label{theo:dagdepth3}
	\dsdts is \NP-hard on DAGs of depth 3, and \W$[2]$-hard when parameterized by~$k$.
\end{theorem}
\begin{proof}
	We reduce from DS. Thus, for an instance $G$ construct the following DAG $G'$:
	
	\underline{Vertices:} For $1\le i\le k$ create vertices $s_i \in S$, $d_i \in D$, $b_i$, and for each vertex $v$ and $0\le i\le k$ create vertices $v_i$. Create vertices $a_1,a_2,\dots, a_7$, with $a_1,a_2,a_3,a_5 \in S$ and $a_1,a_2,a_4,a_7 \in S$.
	
	\underline{Edges:} Add the edges $a_3a_4, a_5a_6, a_6a_4 $ and $ a_5a_7$. For each $1\le i \le k$ add the edge $a_7b_i$. For each $v\in V(G)$ create the edge $v_0a_3$. For each $1\le i\le k$ and $v\in V$ create the edges $s_iv_i, v_id_i, v_ia_1 $ and $v_ib_i$ as well as $v_0d_i$.
	Lastly for each $1\le i\le k$ and each pair $u,v \in V(G)$ with $v\in N[u]$ add the edge $u_0v_i$.
	%\alex{also add $v_0,d_i$ for all $i\le k$ right ?}
	
	For one direction let $\{x_1, x_2,\dots,x_k\}$ be a dominating set for $G$. For each $i=1,2,\dots,k$ move the token from $s_i$ to $u_i$, with $u = x_i$ dominating (all) vertices $d_i$. As $X$ is a dominating set, all $(v_0)_{v\in V}$ are dominated,%\alex{also all $b_i$ right?}
	 thus we may move the token from $a_3$ to $a_4$, followed by a move from $a_5$ to $a_7$. As $a_7$ dominates all $(d_i)_{i\le k}$, we can move the token from $u_1$ to $d_1$, dominating all the $(v_0)_{v\in V}$, enabling all token on $u_{i>1}$ to mote to $d_i$. This conclude the reconfiguration sequence, so $(G',S,D)$ is a positive instance for \dsdts.
	
	For the other direction, first observe that the tokens from $a_1$ and $a_2$ may not move at all, and the tokens from $a_3$ can only move to $a_4$ and the one from $a_5$ only to $a_7$, by reachability. The vertex $a_6$ requires that the token on $a_3$ moves before the token on $a_5$ moves.
	Thus, when $a_3$ moves, both each $(v_0)_{v\in V}$ and $b_{i\le k}$ have to be dominated by some $u_j$ ($j=1,2,\dots,k$). 
	Therefore, the set $X$ of all $u$ in $V$ such that there is a token on some $(u_i)_{1\le i\le k}$ must be a dominating set for $G$.
\end{proof}

\begin{theorem}
	\label{theo:dagdepth2}
	For DAGs of depth two, \dsdts is solvable in linear time.
\end{theorem}
\begin{proof}
	We show that there exists a reconfiguration sequence, exactly if $S\cap D$ dominates all vertices in $V(G)\setminus (S \cup D)$ and there is a matching between $S\setminus D$ and $D \setminus S$.
	
	Obviously, if this condition is met, we can just slide the tokens along the matching. For the other direction, suppose towards a contradiction, that there is a vertex $v\notin S\cup D$ not dominated by $S\cap D$.
	There has to be an $s\in S$ on the layer 1 and $d \in D$ in layer 2 that dominates $v$. However, as there are no connections within one layer, $v$ can neither lay at layer 1 nor at layer 2. Therefore, $S\cap D$ must dominate $G\setminus (S\cup D)$. Then, as tokens in $S \cap D$ cannot move, if there is no matching between $S\setminus D$ and $D\setminus S$, some end-position cannot be reached.
\end{proof}
    % \section{DAGs and treewidth (preliminary)}
    % \subfile{./Sections/DAGs_and_Treewidht.tex}

    % \newpage
    \section{Undirected reconfiguration}\label{sec:undirected-reconfig}

We now turn to our main contribution and look at dominating set reconfiguration for undirected graphs. We study the red blue variant before looking at the implications for the uncolored version. Before that we discuss the properties of a specific gadget that is going to be used throughout the rest of this paper. 

\subsection{The validator gadget}

\begin{figure}[H]
	\includegraphics[width=\linewidth]{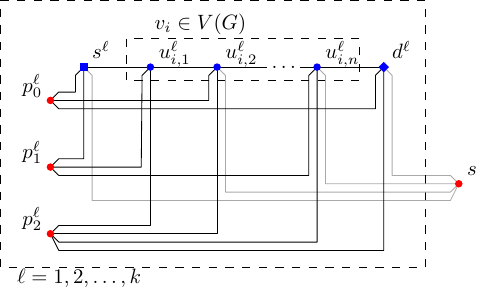}
	\caption{The $\validator(G,k)$ gadget, where $v_i$ dominates $v_2$ and $v_n$ but not $v_1$ in $G$. In this figure we assume that $n\equiv 1 \mod 3$.}
	\label{fig:validator}
\end{figure}

\begin{Gadget}[Validator Gadget]
	\normalfont
	For a graph $G$ with vertices $v_1,\ldots,v_n$ and an integer $k$ the red-blue graph $\validator(G,k)$ together with sets $S$ and $D$ is defined as:
	
	% ...construct the red-blue graph $\validator(G,k)$ together with sets $S$ and $D$ are defined as follows:\jona{I think something went wrong while merging. Choose one (I think in any case we have to mention $S$ and $D$)}
	
	\underline{Blue Paths:} 
	For each $\ell\in\interval{1}{k}$ create $s^\ell \in S$ and $d^\ell \in D$. 
	For each $v_i \in V(G)$ create a path $s^\ell-v^\ell_{i,1}-v^\ell_{i,2}-\dots -v^\ell_{i,n}-d^\ell$. We slightly abuse notations and use~$u^\ell_{i,0}$ instead of $s^\ell$ and $u^\ell_{i,n+1}$ instead of $d^\ell$ (for arbitrary $i$.)
	
	\underline{Red Vertices:}
	Create a red vertex $s$ and for each $\ell\in\interval{1}{k}$, add 3 vertices $p^\ell_{1},p^\ell_{2}$ and~$p^\ell_{3}$.
	For each $\ell\in\interval{1}{k}$ and $i,j \in \interval{1}{n}$, add the edge $(u^\ell_{i,j},s)$ if and only if $(v_i,v_j) \in E(G)$ or~$v_i=v_j$.
	Lastly, for $i$ as before and $j \in\interval{0}{n+1}$ connect $u^\ell_{i,j}$ to $p^\ell_{[j\mod 3]}$ and $p^\ell_{[j +1 \mod 3]}$.
\end{Gadget}
The construction is visualized in \cref{fig:validator}.

% \jona{Do we really want to add these dashes for the path notation? I think then we have to update it everywhere.}
% \alex{I like it here, not sure we need to mark it everywhere, I'll keep it in mind}
% \jona{I like the mod notation, but do we have to add that everywhere? Or do we say $[1 \mod 3] = 1$ (and so on)}
% \alex{We should comment on it in the prelims, both reminder and equiv classes}
% \jona{Ok}

\begin{lemma}\label{lem_validator-dominates_s}
	Let $G$ be a graph, $k$ an integer, and $I= \{i_1,\ldots,i_k\}\subseteq \interval{1}{n}$ a set of $k$ indices.
	Then $\{v_{i_1},\ldots,v_{i_k}\}$ is a dominating set in $G$ if and only if for every $j\le n$ there is an $\ell\le k$ such that $(u^\ell_{i_\ell,j},s)$ is an edge of $\validator(G,k)$
\end{lemma}
\begin{proof}
	Assume $\{v_{i_1},\ldots,v_{i_k}\}$ is a dominating set of $G$, let $j\le |V|$ be any index of $G$, there must be a vertex $v_{i_\ell}$ that dominates $v_j$ hence in $\validator(G,k)$, $u^\ell_{i_\ell,j}$ is adjacent to $s$.

	For the other direction, let $v_j$ be any vertex of $G$, by assumption there is an $\ell\le k$ such that $u^\ell_{i_\ell,j}$ is adjacent to $s$ in $\validator(G,k)$, hence $v_{i_\ell}$ dominates $v_j$ in $G$, so $\{v_{i_1},\ldots,v_{i_k}\}$ is a dominating set of $G$.
\end{proof}

We will now define two properties that let us transform \rbdsts instances into regular \dsts instances. \Cref{prop-1blue} is needed to "trap" tokens in a blue component, while \cref{prop-pw-dec} is used th maintain tight bounds on the path and tree-decomposition.

\begin{property}[Single token per component]\label{prop-1blue}
	An instance $(G,S,D)$ of \rbdsts is said to have \cref{prop-1blue} if and only if every blue connected component of $G$ contains exactly one vertex of $S$ (equivalently one vertex of $D$).
\end{property}

\begin{property}[Blue components in series]\label{prop-pw-dec}
	A path (resp. tree) decomposition of a Red-Blue graph $G$ is said to have \cref{prop-pw-dec} if and only if there is a collection of disjoint subpaths (resp. subtrees) ${\cal B}$ and a one to one matching between $\cal B$ and the collection of blue connected components $\cal C$ of $G$, such that for every matched pair $C\in{\cal C}, B\in{\cal B}$, we have that every vertex of $C$ appears in at least one bag of $B$.
\end{property}
% Both \cref{prop-1blue} and \cref{prop-pw-dec} generalize to directed graph by looking at \textcolor{blue}{weakly} connected components\jona{what do you mean by connected component? Also blue? I think for complete components, we don't need prop 2, as we can allways some them to be at different positions in the decomp}\alex{forgot weakly, but yes lets remove it all} and path decomposition of the underlying undirected graphs.\todo{remove paragraph}

\subsection{Red-Blue Domination}
\begin{theorem}
	\label{theo:rbdstsW2}
	\rbdsts is W[2]-hard on undirected graphs for the parameter $k+\pw+\iota$. Even on graphs with \cref{prop-1blue} together with a path decomposition having \cref{prop-pw-dec} of width~$3k+4$.
\end{theorem}

\begin{proof}
	We reduce from dominating set on undirected graphs. Let $G$ be any undirected graph and $k$ an integer. We build a graph $H$ composed of $\validator(G,k)$, and a path $P$ of $n+2$ blue vertices $h_0\ldots h_{n+1}$ where $h_0\in S$ and $h_{n+1}\in D$.
	In addition to the edges within the path $P$ and validator gadget, for every $0\le j\le n+1$ we have the edges between $h_j$ and $p^\ell_{j'}$ if and only if $j'\equiv j$, or $j'\equiv j+2\mod 3$. We soon show that $(H,S,D,\iota=1)$ is a positive instance of \rbdsts if and only if $G$ has a dominating set of size $k$.
	But before that we prove that:
	\begin{claim}
		$(H,S,D)$ has \cref{prop-1blue}, and there is a path decomposition of $G$ with \cref{prop-pw-dec} and width $3k+4$.
	\end{claim}
	\begin{claimproof}
		The red vertices $p^\ell_j$ and $s$ split the blue vertices into $k+1$ blue connected component. The blue path $h_0-\ldots-h_{n+1}$ is one and only has  $h_0$ in $S$.
		For every $\ell$, we have that $\{s^\ell,d^\ell\}\cup\bigcup_{i,j}u^\ell_{i,j}$ is one connected component with only $s^\ell$ in $S$. For the path decomposition each blue connected component has pathwidth at most $3$ (by adding $s^\ell,d^\ell$ in each bag and then taking the $n$ rows one after the other, two vertices at a time). Concatenating these decompositions in any order and adding the $3k+1$ red vertices in every component yields the desired  path decomposition of~$H$.
	\end{claimproof}

	\underline{Description of the reconfiguration sequence:}
	Assume first the existence of a dominating set $(v_{i_\ell})_{\ell\in I}$ for $G$. First for every $\ell\le k$ we move every from $s_\ell$ to $u^\ell_{i_\ell,1}$. These moves can be preformed because $h_0$ dominates $p_0^\ell$ and $p_2^\ell$, and $u^m_{i_m,1}$ dominated $s$, where $m\le k$ is such that $v_{i_m}$ dominates $v_1$ in $G$. Then $h_0$ can move to $h_1$. 

	We continue by repetitively moving every $u^\ell_{i_\ell,j}$ to $u^\ell_{i_\ell,j+1}$ then $h_j$ to $h_{j+1}$. Every time the first token that move is the one for which $u^\ell_{i_\ell,j+1}$ dominates $s$. We show in the next claim that such token always exists.

	\begin{claim}\label{claim-exist-reconfig}
		This is a valid reconfiguration sequence.
	\end{claim}
	\begin{claimproof}
		Let $j\in\interval{0}{n+1}$ assume the tokens are on position $h_j$ and $u^\ell_{i_\ell,j}$ for every $\ell$. Then as only the $p^\ell_{j'}$ for $j'\equiv j+1\mod 3$ are not dominated by $h_j$, and since both $u^\ell_{i_\ell,j}$ and $u^\ell_{i_\ell,j+1}$ dominate $p^\ell_{j'}$ we can move to $u^\ell_{i_\ell,j+1}$. Furthermore, by \cref{lem_validator-dominates_s} there is one of the $\ell\le k$ such that $u^\ell_{i_\ell,j+1}$ dominates $s$.

		After that, only the $p^\ell_{j'}$ for $j'\equiv j\mod 3$ are not dominated by the token in the validator gadget. And since both $h_j$ and $h_{j+1}$ dominate every $p^\ell_{j'}$ for this value of $j'$ (as $j+1+2\equiv j\mod 3$), we can move from $h_j$ to $h_{j+1}$.
	\end{claimproof}
	By considering $d^\ell$ as $u^\ell_{i,n+1}$, this concludes the first direction of the proof.
	Assume now that $(H,S,D,\iota=1)$ is a positive instance of \rbdsts.
	Before $h_0$ moves $h_1$, no token can be on some $u^\ell_{i,2}$ as some $p^\ell_1$ would not be dominated. Furthermore, when $h_0$ reaches $h_1$ no token can still be on $s^\ell$ as $p_2^\ell$ would be dominated. 
	Thus, when $h_0$ moves to $h_1$, every token must be on some $u^\ell_{i_\ell,1}$. We fix $I=\{i_1,\ldots,i_\ell\}$. \textit{We say that token $\ell$ chose the path $i_\ell$}. 
	\footnote{An astute reader would notice that, if not for the bounded iteration, the token on the $h_j$ could move back and forth. We would then define $I$ for the last time that the token leaves $h_0$. The following lemma holds when the token moves to $h_j$ either from $h_{j-1}$ or $h_{j+1}$. This is highlighted in the proof of \cref{theo:rbdstsW2-fix-pw}.}\linebreak
	We then claim that the tokens remain synchronized.
	\begin{claim}
		For every $j$, when the token on $h_{j-1}$ moves to $h_j$, for every $\ell\le k$, the $\ell$th token (the token starting on $s^\ell$) of the validator gadget is on $u^\ell_{i_\ell,j}$. 
	\end{claim}
	\begin{claimproof}
		By induction on $j$, when the token in $P$ entered $h_{j-1}$ every token $\ell$ was on $u^\ell_{i_\ell,j-1}$. Assume for clarity and wlog that $j-1\equiv 0\mod 3$. Then $h_{j-1}$ dominates every~$p^\ell_0$ and~$p^\ell_2$. So each token on $u^\ell_{i_\ell,j-1}$ is free to move to $u^\ell_{i_\ell,j}$ but cannot move back to $u^\ell_{i_\ell,j-2}$ nor to~$u^\ell_{i_\ell,j+2}$ as $p^\ell_1$ would be undominated.
		Furthermore, when $h_{j-1}$ moves to $h_j$, no token can still be on $u^\ell_{i_\ell,j-1}$ as $p^\ell_2$ would be undominated. Therefore, the tokens are on $u^\ell_{i_\ell,j}$ for each $\ell\le k$.
	\end{claimproof}
	Now note that for every $j\le n$, when the token enters $h_j$, there must be a token on $u^\ell_{i_\ell,j}$ for some $\ell\le k$ that dominates $s$ so by \cref{lem_validator-dominates_s}, $(v_i)_{i\in I}$ is a dominating set for $G$.
\end{proof}

In \cref{theo:rbdstsW2} the pathwidth is a function of $k$. Not bounding the iteration parameter enables shrinking the pathwidth and treewidth numbers to a constant.
\begin{theorem}
	\label{theo:rbdstsW2-fix-pw}
	\rbdsts is W[2]-hard for the parameter $k$ on undirected graphs with \cref{prop-1blue}, 
	a path decomposition and a tree decomposition, both with \cref{prop-pw-dec} of width $10$ and $6$ respectively.
\end{theorem}

\begin{proof}
	We reduce again from dominating set on undirected graphs. Let $G$ be any undirected graph and $k$ an integer. We build a graph $H$ composed of $\validator(G,k)$, and a cycle of~$3$ blue vertices $h_0,h_1,h_2$ where $h_0\in S$ and $h_{j}\in D$ for $j\equiv n+1\mod 3$.
	In addition to the edges within the cycle and validator gadget, for every $0\le j\le n+1$ we have the edges between $h_j$ and $p^\ell_{j'}$ if and only if $j'\equiv j$, or $j'\equiv j+2\mod 3$.
	We call the token that starts on $h_0$ the {\em cycle token} and the others the {\em gadget tokens}.
	We soon show that $(H,S,D)$ is a positive instance of \rbdsts if and only if $G$ has a dominating set of size $k$.
	But first we prove that:
	\begin{claim}
		$(H,S,D)$ has \cref{prop-1blue}, treewidth $6$, and a path decomposition with \cref{prop-pw-dec} of width $10$.
	\end{claim}
	\begin{claimproof}
		As in the proof of \cref{theo:rbdstsW2} the red vertices split the graph into $k+1$ blue connected components, each with one vertex in $S$.

		For the path decomposition, for each $\ell\le k$ the graph with $s^\ell,d^\ell$, the $p_j^\ell$, and the $u_{i,j}^\ell$ has pathwidth $6$. Adding the 3 blue vertices $h_0$, $h_1$, $h_2$ and $s$ into each bag yields pathwidth $10$. Concatenating these decompositions for each $\ell$ and adding a bag at the beginning containing only $h_0$, $h_1$, and $h_2$
		yields a path decomposition with \cref{prop-pw-dec}.

		Concerning treewidth, we can build the root containing the 3-cycle and $s$. Then for each~$\ell$ we build a child containing the same vertices as the root and adding $p^\ell_{0}, p^\ell_{1} $ and $ p^\ell_{2}$. Each such node gets a child where we remove the 3-cycle and add $s^\ell$ and $d^\ell$. 
		We then branch into $n$ many children for each $i\le n$ to take the decomposition of the path $(u^\ell_{i,j})_{j\le n}$ (keeping $s$, $d^\ell$ and $p^\ell_{0}, p^\ell_{1}$ and $ p^\ell_{2}$ in each bag). This yields a maximal bag with 7 vertices, hence treewidth $6$. To see that this decomposition has \cref{prop-pw-dec}, pair the 3-cycle component with the root of the decomposition; and pair the $\ell$th blue connected component of the validator gadget with the subtree rooted on the bag where $s^\ell,d^\ell$ are introduced.
	\end{claimproof}

	Concerning the rest of the proof. If $G$ has a dominating set of size $k$, the existence of a reconfiguration sequence is a straightforward adjustment of \cref{claim-exist-reconfig}, by replacing~$h_j$ by~$h_{j'}$ such that $j'\equiv j\mod 3$.
	The rest of the proof is also somewhat similar except we now cannot deduce the column $j$ of the tokens in the $\validator$ gadget from the position of the token in $h_0,h_1,h_2$ we have to remember how many laps have occurred.
	
	To this end, we call a move from $h_j$ to $h_{j+1}$ (with $2+1=0$) a {\em forward} move and a move from $h_{j+1}$ to $h_j$ a {\em backward} move.

	\begin{claim}\label{claim:backward-forward}
		Let $a$ be the number of forward moves and $b$ the number of backward moves. At all time $a\ge b-1$, and, right after a forward (resp. backward) move, the $\ell$th token of the $\validator$ gadget is on some $u^\ell_{i,j}$ where $j=a-b$ (resp $j=a-b+1$). 
	\end{claim}
	\begin{claimproof}
		The proof is done by induction on $a+b$ and holds for $a=b=0$. If $a-b=-1$ then by induction the last move was a backward move and the gadget tokens are on $u^\ell_{i,0}$ i.e.~$s^\ell$. Furthermore, the cycle token in on $h_2$ and does not dominate $p^\ell_0$, so the gadget tokens are locked and a backward move (to $h_1$) would leave $p_2^\ell$ undominated. So $a$ must remains at least $b-1$.

		Now assume that $a\ge b$ and the previous move was a forward move. So all gadget token are on some $u^\ell_{i,j}$ with $j=a-b$. Assume wlog that $j\equiv 0 \mod 3$. The cycle token can immediately revert its forward move and the property would be satisfied. But before making another forward move (to $h_1$), every gadget vertex need to move toward $u^\ell_{i,j+1}$ as otherwise~$p^\ell_2$ would be undominated.

		Finally, assume $a\ge b$ and the previous move was a backward move. This is analogous to the previous case, where the cycle token can immediately make a forward move, or every gadget vertex goes to $u^\ell_{i,j-1}$ before the cycle token makes another backward move.
	\end{claimproof}

	To conclude the proof, note that when the reconfiguration sequence ends there must have been $n+2$ more forward moves than backward ones. We look at the last time the number of backward and forward moves are equals, and similarly to the proof of \cref{theo:rbdstsW2} after the next forward move, every token must be on some $u^\ell_{i_\ell,1}$. We fix $I=\{i_1,\ldots,i_\ell\}$. We then conclude with \cref{lem_validator-dominates_s} that $(v_i)_{i\in I}$ is a dominating set for $G$. 
\end{proof}
    \subsection{Consequences for classical dominating set reconfiguration}

\begin{lemma}
	\label{lem:red-blue-to-dsr-undirected}
	For any \rbdsts instance $(G, S, D,\iota)$ with \cref{prop-1blue} and a path (resp.~tree) decomposition of width $p$ with \cref{prop-pw-dec}. We can build in time $O(\|G\|)$ a graph $G'$ and sets~$S', D'$, such that $\pw(G') \le p+3$ (resp.~$\tw(G')\le p+3$), $|S'|=|D'|=|S|+1$,
	and $(G',S',D',\iota)$ is a positive instance of \dsts if and only if $(G,S,D,\iota)$ is a positive instance of \rbdsts.
\end{lemma}

\begin{proof}
	First, start with $(G', S', D') = (G, S, D)$. Second add two adjacent vertices $d$ and $d'$; add $d$ to both $D'$ and $S'$ and connect $d$ to every blue vertex of $G'$.

	Lastly, for every blue connected component $C$ add two independent vertices $d_C$, and $d'_C$, both adjacent to every vertex of $C$.
	
	Assume that $(G,S,D,\iota)$ is a positive instance of \rbdsts. Then perform the same moves in $G'$. The new vertices $d_C,d'_C$ are dominated by the token that remains on $C$ (which exists by \cref{prop-1blue}), the vertices that were red in $G'$ are dominated throughout the reconfiguration by definition, and the previously blue vertices are all dominated by the new vertex $d$.
	
	For the other direction assume now that $(G',S',D',\iota)$ is a positive instance of \dsts. Note that the token on $d$ cannot move, as $d'$ would be undominated.
	Similarly, for every blue connected component $C$, the only token that starts in $C$ (which exists by \cref{prop-1blue}) must remains in $C$ as either $d_C$ or $d'_C$ would be undominated.
	Last while $d$ dominates every vertex that is blue in $G$, the red vertices must be dominated by the tokens placed on the blue vertices.
	Therefore, every move that is preformed on $(G',S',D',\iota)$ can also be performed on~$(G,S,D,\iota).$
	
	For the structural parameter, adding $d$ to every bag and adding one bag with $d$ and $d'$ increases the pathwidth (resp.~treewidth) of the instance by only 1.
	For the vertices $d_C,d'_C$, \cref{prop-pw-dec} provides the set of bags to which they can be added without interfering, increasing further the pathwidth (resp.~treewidth) by only 2.
\end{proof}

The next statement, which is our main result, is now a direct consequence of \cref{theo:rbdstsW2} and \cref{lem:red-blue-to-dsr-undirected}. 

\begin{theorem}
	\label{theo:dstsW2}
	\dsts is W[2]-hard on undirected graphs parametrized by $k+\pw+\iota$.
\end{theorem}

Similarly to the red-blue variant, not including $\iota$ in the parameters enables to fix the treewidth and pathwidth of the instances by applying \cref{lem:red-blue-to-dsr-undirected} to \cref{theo:rbdstsW2-fix-pw} instead of \cref{theo:rbdstsW2}.

\begin{theorem}
	\label{New2}
	\dsts is W[2]-hard when parametrized by $k$ on undirected graphs with pathwidth $13$ and treewidth $10$.
	% \jona{I think it also has to be 10 here. Maybe we can shrink by using a two cycle(essentially just as in the other theorem (But is it at this point really worth writing?))}\alex{lets keep 10, w'ill see after submitting whether we improve}
\end{theorem}

    \section{Reconfiguration on Directed Graphs}\label{sec:directedReconfig}

We now turn to reconfiguration on directed graph, where tokens can only follow the edge direction. While it is known that \isdts is W[1]-hard on DAGs (parametrized by $k$, the size of the independent set), it becomes FPT when the treewidth of the instance is added to the parameter. We show that this is not the case for \dsdts. As in \cref{sec:undirected-reconfig}, we start with the analysis of red-blue dominating set reconfiguration, this time for {\em directed} token sliding \rbdsdts. 
\subsection{Red Blue Directed Reconfiguration}

\begin{theorem}
	\label{theo:rbdsdtsDAGW2}
	\rbdsdts is W[2]-hard for DAGs for the parameter $k+\pw$. Even on instances with a path decomposition of width $3k+4$.
\end{theorem}

Relaxing the DAG constraints yields better bounds

\begin{theorem}
	\label{theo:rbdsdtsW2hardfixtw}
	\rbdsdts is W[2]-hard for directed graphs the parameter $k$ for fixed $\tw = 6$, $\pw=9$ and $\dfvs = 1$.
\end{theorem}

The proof are straightforward adaptations of \cref{theo:rbdstsW2,theo:rbdstsW2-fix-pw}. First, in the definition of the validator gadget, orient the edges from $u^\ell_{i,j}$ to $u^\ell_{i,j+1}$ and edges from blue vertices toward red vertices. 
Second in the reductions make a directed path $h_0\rightarrow\ldots\rightarrow h_{n+1}$, (or a directed $3$-cycle $h_0\rightarrow h_1\rightarrow h_2\rightarrow h_0$).
The rest of the proofs follow and require even less details as the tokens cannot ``go back''. More formally, in the proof of \cref{claim:backward-forward} there can be no backward moves.  

    \subsection{Consequences for directed reconfiguration of dominating set}

The fact that token movements are more constrained in directed graphs enables us to prove the following lemma, analogous to \cref{lem:red-blue-to-dsr-undirected} with slightly better bounds.
% while maintaining the number of directed cycle.

% \begin{lemma}
% 	\label{red-blue-to-dsr-directed}
% 	For any \rbdsdts instance $(G, S, D)$ with \cref{prop-1blue} and a path (resp.~tree) decomposition of width $p$ with \cref{prop-pw-dec}. We can build in time $O(\|G\|)$ a graph $G'$ and sets~$S', D'$, such that $\pw(G') \le p+1$ (resp.~$\tw(G')\le p+1$), $|S'|=|D'|=|S|+1$, $\dfvs(G')=\dfvs(G)$\alex{check}
% 	and $(G',S',D')$ is a positive instance of \dsdts if and only if $(G,S,D)$ is a positive instance of \rbdsdts.
% \end{lemma}
Here again we can assume wlog that all edges in $G$ between red and blue vertices start in blue and end in red, and that there are no edges between two red vertices

\begin{lemma}
	\label{red-blue-to-dsr-directed}
	For any \rbdsdts instance $(G, S, D)$. We can build in time $O(\|G\|)$ a graph $G'$ and sets~$S', D'$, such that $\pw(G') = \pw(G)+1$, $\tw(G')=\tw(G)+1$, $|S'|=|D'|=|S|+1$, $\dfvs(G')=\dfvs(G)$ and $(G',S',D')$ is a positive instance of \dsdts if and only if $(G,S,D)$ is a positive instance of \rbdsdts.
\end{lemma}
\begin{proof}
	The only thing we do is that we add a vertex $d$ in $S\cap D$ and connect it toward every blue vertex of~$G$. 
	Clearly every structural consideration is matched.

	Assume $(G,S,D)$ is a positive instance of \rbdsdts. As we did not change the orientation of the edges, we then perform the same moves in $G'$. At all time the vertices that are red in $G$ are dominated by the tokens, and the blue vertices (in $G$) are dominated by the token in $d$.

	For the other direction, assume now that $(G',S',D')$ is a positive instance of $\dsdts$.
	First note that no token can reach $d$, so the token that starts in $d$ cannot move at all. Furthermore, the orientation of edges ensures, that no token who leave its blue connected component can reach a destination.
	So every move can be replicated in $G$.
	Lastly, the tokens that are not in $d$ must dominate every vertex that is not dominated by $d$, so in $G$ every red vertex is dominated throughout the reconfiguration sequence.
\end{proof}

This applied to the hard instances of \cref{theo:rbdsdtsDAGW2,theo:rbdsdtsW2hardfixtw} yields the following results: 

\begin{theorem}
	\label{theo:dsdtsdaghard}
	\dsdts is W[2]-hard for DAGs when parametrized by $k+\pw$.
\end{theorem}

Relaxing the DAG constraints to small directed feedback vertex set enables to get hardness already of directed graphs of fix pathwidth and treewidth.

\begin{theorem}
	\label{theo:dsdtsdagharddfvs}
	\dsdts is W[2]-hard when parametrized by $k$ for directed graphs with treewidth $7$, pathwidth $11$, and $\dfvs = 1$.
\end{theorem}

% \Alex{Sketch for both}
% use \cref{red-blue-to-dsr-directed} reformulated a bit. Keeps the number of directed cycle so DAGs remains DAGs, and DFVS stay 1

    % \newpage
    \section{NP hardness for small treewidth and pathwidth}\label{sec:np-best-bounds}
Not the validator gadget to yields even better bounds but blows up $k$.

At the cost of blowing up the number of token, it is possible to get a better lemma than \cref{lem:red-blue-to-dsr-undirected}, meaning that the treewidth do not increase when transforming an \rbdsdts instance to a \dsdts one.

\begin{lemma}
	\label{red-blue-to-dsr-directed-tw-preserved}
	For a \rbdsdts instance $(G, S, D)$ we can build a \dsdts instance $(G', S', D')$ with $\tw(G') = \tw(G) > 0$ and $\pw(G') = \pw(G) +1 $, such that $(G', S', D')$ is a positive instance if and only if $(G, S, D)$ is. Further if $G$ is a DAG so is $G'$.
\end{lemma}
\begin{proof}
	We build $G'$ from $G$ by creating one new vertex $v'$ for every blue vertex $v$ of $G$, and making a directed edge from $v'$ to $v$. Again we assume that the edges between blue and red vertices are directed from blue to red. We set $S'= S \cup \{v'\mid v \textrm{ is blue vertex of } G\}$ and $D'=D\cup \{v'\mid v \textrm{ is blue vertex of } G\}$.
	
	The tokens on all vertices $v'$ of $G'$ may never leave, as by reachability then no vertex can reach the destination of $v'\in D'$. It follows, that all vertices $v\in V(G')$ that were blue vertices in $G$ remain dominated regardless of the position of the other tokens. Further, by reachability, no token can ever be on a vertex that was red in $G$, as this token can then never reach a destination. 
	Thus, $(G, S, D)$ is a positive instance if $(G', S', D')$, is and vise versa.
	
	We can build a tree decomposition of equal width as follows. We start with the tree decomposition for $G$. Now for all blue vertices $v\in V(G)$ let $x_v$ be a node with $v\in\bag(x_v)$. We add a child $x_v'$ of $x_v$ and set $\bag(x_v') = \{v,v'\}$. 
	
	For the path decomposition, we add $v'$ to one bag that also contains $v$. We duplicate bags if necessary to ensure that only one vertex gets added per bag. 
\end{proof}

We now proof that \rbdsdts is hard even if the treewidth is fixed to a small value.

\begin{Gadget}[Variation of Checking Gadget]
	\label{gadget:variation-checking}
	\normalfont 
	Start with the $\validator(G, k)$ gadget and modify it as follows. For all $\ell$ in $\interval{1}{k}$ and $i$ and $j$ in $\interval{1}{n}$ remove all edges between $u^\ell_{i,j}$ and $p^\ell_{1}, p^\ell_{2}$ and $p^\ell_{3}$. Further remove all vertices $p^\ell_{3}$.

	For each $i,j$ and $\ell$ subdivide the edge 
	$u^\ell_{i,j} u^\ell_{i,j+1}$; 
	let $u'^\ell_{i,j}$ be the new blue vertex in between. 
	Further, connect $u'^\ell_{i,j} $ to $s$ if and only if $u^\ell_{i,j}$ is.
	
	% \st{For each $\ell\in\interval{1}{k}$ connect $p^\ell_{1}$,$p^\ell_{2}$ to $s_i$ and $t_i$.}
	For each $\ell\in\interval{1}{k}$ connect $s_i^\ell$ and $d_i^\ell$ to $p^\ell_{1}$,$p^\ell_{2}$.
	
	For each $\ell\in\interval{1}{k}$ 
	and for each $j,i\in\interval{1}{n}$ 
	connect $u'^\ell_{i,j}$ to both $p^\ell_{1}$, and $p^\ell_{2}$.

	Finally, if $j \equiv 1 \mod 2$ then connect $u^\ell_{i,j}$ to $p^\ell_{1}$, 
	and if $j \equiv 0 \mod 2$ connect $u^\ell_{i,j}$ to $p^\ell_{2}$.
\end{Gadget}

\begin{Gadget}[Clock with Battery]
	\label{gadget:clock-with-battery}
	\normalfont
	Create blue vertices $h_1$ and $h_2$, with $h_2 \in S$.
	Further, let $h_1 \in D$ if $n$ even, and $h_2\in D$ otherwise.
		
 	For $j\in\interval{1}{n-1}$, create blue vertices $z_j,y_j$, red vertex $b_j$ and edges $z_jb_j$ and $y_jb_j$. 
	Finally, connect $h_1$ and $h_2$ to each $z_j$ and each $y_j$ to both $h_1$ and $h_2$.
	% If $j = 1 \mod 2$ connect $h_{1}$ to $z_j$ and  $y_i$ to $h_{2}$. Add $z_j$ to $D$ and $y_j$ to $S$.
 	% Further if $j = 0 \mod 2$ connect $h_{2}$ to $z_j$ and  $y_i$ to $h_{1}$.
\end{Gadget}

\begin{theorem}
	\label{theo:rbdsdtsNPhardfixTW}
	\rbdsdts is \NP-hard for DAGs with fixed \pw = 8 and \tw = 5.
\end{theorem}
\begin{proof}
	We show by reduction from dominating set.
	For an instance $G,k$ of DS, first build the gadgets form \cref{gadget:variation-checking} and \cref{gadget:clock-with-battery}. 
	Connect them by adding edges from $h_1$ (of \cref{gadget:clock-with-battery}) to all $p_{1}^\ell$ ( of \cref{gadget:variation-checking} ), $h_2$ to all $p_{2}^\ell$ for all $\ell\in \interval{1}{k}$. Let $(G', S, D)$ be the resulting instance. We refer to the path $ u^\ell_{i,1}u'^\ell_{i,1}u^\ell_{i,2}u'^\ell_{i,1}\dots u^\ell_{i,n}d^\ell$ as $P_i^\ell$.
	
	\begin{claim}
		$G'$ has a tree-decomposition of size 5.
	\end{claim}
	\begin{claimproof}
		We can construct a tree-decomposition as follows. Let the root have the bag  $\{h_1,h_2, s\}$. 
		For each $j\in \interval{1}{k}$ add node below root with bag $\{h_1,h_2,z_j,y_j\}$ and below this a node with $\{z_j,y_j, b_j\}$.
		
		For each $\ell\in\interval{1}{k}$ add node $x^\ell_{1}$ with bag containing $h_1,h_2,p^\ell_{,1},p^\ell_{,2}$ and $q$ below the root. 
		Further add a node $x^\ell_{2}$ with bag $p^\ell_1,p^\ell_2,s^\ell,d^\ell,s$ below $x^\ell_{1}$ respectively.
		For each $\ell$ as before and $i\in\interval{1}{k}$ add a child of $x^\ell_{2}$ with bag containing $p^\ell_{1},p^\ell_{2},s^\ell,d^\ell,s$ and $u^\ell_{i,j}$ . 
		Below that add a path with $j\in\interval{2}{n}$ of nodes with bags $\{p^\ell_{1},p^\ell_{2},d^\ell,s,u^\ell_{i,j-1},u^\ell_{i,j}\}$.

		The largest bag contains $6$ vertices, hence the treewidth is $5$.
	\end{claimproof}
	
	\begin{claim}
		$G'$ has a path-decomposition of size 8.
	\end{claim}
	\begin{claimproof}
		A path-decomposition may be constructed as follows.
		Looping through $\ell\in\interval{1}{k}$, $i\in\interval{1}{n}$ and $j\in\interval{1}{n}$ in this order, make a bag $\{s^\ell,d^\ell, u^\ell_{i,j},u^\ell_{i,j-1},p^\ell_1,p^\ell_2\}$ and attached it to the previous bag.
		For each $j=1,\dots,n-1$ make a bag containing $z_j,b_j $ and $y_j$.
		Last, add to each bag $q, h_1 $ and $h_2$, yielding $9$ vertices per bag.
	\end{claimproof}

	\begin{claim}
		If $(G',S,D)$ is a positive instance for \dsdts then $G,k$ has a dominating set of size $k$.
	\end{claim}
	\begin{claimproof}
		Take any reconfiguration sequence $(\alpha_a)_{a\in A}$ certifying that $(G',S,D)$ is a positive instance.
		
		In this sequence in each step there cannot be a token on both $h_1$ and $h_2$. No token can enter from outside the subgraph of \cref{gadget:clock-with-battery} and all $b_j$ ($j \in \interval{1}{n-1}$) need to be dominated. This leaves at most one spare token which may be on $h_1$ or $h_2$.
		
		Take pairs of indices $(i,j)$ such that over all configurations $\alpha_i,\alpha_{i+1},\dots,\alpha_{j}$ a token is either only on $h_1$ or only on $h_2$ with $i$ being minimal and $j$ maximal. There are at most $n$ such pairs.
		
		This has consequences for the subgraph constructed in \cref{gadget:variation-checking}. First a token starting on some $s^\ell$ has to reach $d^\ell$ following a path $P_i^\ell$ for a fixed $i$, by construction. As the graph is a DAG it may never go backwards on this path. 
		
		As the connection of $u^\ell_{i,j}$ (for all $\ell$ and $i$) oscillates between having a connection ot $p^\ell_{1}$ or $p^\ell_{2}$, a token some $s^\ell$ may thus only move two vertices along the path $P_i$ in each of these segments between two indices.
		As the path has $2n$ vertices it needs to move exactly two vertices in each segment. Arguing analogous \cref{theo:rbdstsW2} there exists a dominating set of size $k$ in $G$ as the vertex $s$ of \cref{gadget:variation-checking} as to be dominated in each configuration.
	\end{claimproof}
	
	\begin{claim}
		If $G$ has a dominating set of size $k$, then there is a reconfiguration sequence for $(G',S,D)$.
	\end{claim}
	\begin{claimproof}
		Let $X$ be a dominating set of size $k$ for $G$. Without loss of generality we fix some order of the vertices in it.
		Move the token form $s^\ell$ on some path $P_i^\ell$ such that $v_i$ is the $\ell$th vertex of $X$ move the token as far as possible on that path (to $u'\ell_{i,1}$). Do that for every $\ell$.
		
		This makes it possible to move the token on $h_2$ to $z_1$, which makes it possible to move the token from $y_1$ onto $h_1$.
		Afterwards we move the tokens on $P_i^\ell$ (for all $i$ and $\ell$) again by two vertices. Then again move the tokens on the graph from \cref{gadget:clock-with-battery} such that there is a token on $h_2$.
		Repeat this until all tokens are at $d^\ell$ for all $\ell$ in the subgraph of \cref{gadget:variation-checking}.
		The destination configuration is also reached for the subgraph of \cref{gadget:clock-with-battery}, as this requires exactly $n-1$ changes of the clock. Furthermore, the vertex $s$ is dominated through the reconfiguration as $X$ is a dominating set analogous to \cref{theo:rbdstsW2}.
	\end{claimproof}
	This then concludes the proof of \cref{theo:rbdsdtsNPhardfixTW}
\end{proof}

\begin{theorem}
	\label{theo:dsdtsNPhardfixTW}
	\dsdts is \NP-hard for DAGs with fixed \pw = 9 and \tw = 5.
\end{theorem}
\begin{proof}
	This follows directly from \cref{theo:rbdsdtsNPhardfixTW} and \cref{red-blue-to-dsr-directed-tw-preserved}.
\end{proof}

    \bibliography{ref.bib}
\end{document}